\documentclass[a4paper,11pt]{article}
\usepackage{latexsym}
\author{Wen-Xiu~Ma\footnote{Email: mawx@cityu.edu.hk}\\
Department of Mathematics, City University of Hong Kong,\\
Kowloon, Hong Kong}
\title{Extension of Hereditary Symmetry Operators}
\date{\nonumber}

\begin{document}
\maketitle

\newcommand{\eqnsection}{
   \renewcommand{\theequation}{\thesection.\arabic{equation}}
   \makeatletter
   \csname $addtoreset\endcsname
   \makeatother}
\eqnsection

\newtheorem{thm}{Theorem}[section]
\newtheorem{Le}{Lemma}[section]
\newtheorem{defi}{Definition}[section]
\newcommand{\R}{\mbox{\rm I \hspace{-0.9em} R}}

\def\be{\begin{equation}}
\def\ee{\end{equation}}
\def\bea{\begin{eqnarray}}
\def\eea{\end{eqnarray}}
\def\ba{\begin{array}}
\def\ea{\end{array}}
\def\la {\lambda}
\def \part {\partial}
\def \al {\alpha}
\def \de {\delta}


\begin{abstract}
Two models of candidates for  
hereditary symmetry  operators are proposed and thus many nonlinear systems
of evolution equations possessing infinitely many commutative symmetries
may be generated. Some concrete structures of hereditary symmetry  operators
are carefully analyzed on the base of the resulting general conditions
and several corresponding nonlinear systems 
are explicitly given out as illustrative examples. 
\end{abstract}

\section{Introduction}
\setcounter{equation}{0}

An application of Lax pairs is a well-known way to construct
nonlinear integrable systems. Most integrable systems, such as 
the KdV, the NLS, the KP and the Davey-Stewartson
equations, can be derived through
appropriate Lax pairs (see for example \cite{Konopelchenko}).
There are also some other ways to construct nonlinear 
integrable systems, for example, by bi-Hamiltonian formulation
\cite{Magri,GelfandD}
and by hereditary symmetry  operators \cite{FuchssteinerF,FokasF} 
etc. 
  
Of course, integrable systems generated by different methods 
have different integrable properties. In general, the method of Lax pair  
produces S-integrable systems and 
the methods of bi-Hamiltonian formulation and hereditary symmetry  operators 
produce nonlinear systems possessing infinitely many symmetries
and/or infinitely many conserved densities. 
There has already been a lot of investigation on the method of Lax pair
(see for example \cite{AblowitzC}) 
and the method of bi-Hamiltonian formulation (see for example 
\cite{MagriMT,SantiniF,DorfmanF}).
So far, however, there has been little discussion about the method of 
hereditary symmetry  operators.

This paper will focus on the construction of hereditary symmetry  operators and
their related nonlinear systems. The resulting nonlinear systems 
have infinitely many commutative symmetries. Some of such systems may be found
in Refs. \cite{Fokas,MikhailovSS,ZakharovK,Lu}. However by our idea,
we can easily construct as many such systems as we want.
To achieve our aim, 
we first discuss the structure of hereditary symmetry  operators by 
examining two models of candidates for hereditary symmetry  operators,
and then exhibit some concrete examples of hereditary symmetry  operators
including relevant nonlinear systems.

Let $u$ be a dependent variable $u=(u^1,\cdots,u^q)^T$, where 
$u^i,\ 1\le i\le q,$ 
depend on the spatial variable $x$ and on the temporal variable $t$. We
use ${\cal A} ^q$ to denote the space of $q$ dimensional column 
vector functions depending on $u$ itself 
and its derivatives with respect to the spatial variable $x$ (possibly 
a vector). Sometimes we write this space as ${\cal A}^q(u)$ 
in order to show the 
dependent variable $u$.

\begin{defi} Let $K,S\in {\cal A}^q$ and 
 $\Phi (u):{\cal A}^q\to {\cal A}^q$. Then the Gateaux derivatives
of $K$ and $\Phi$ with respect to $u$ at
 the direction $S$ are defined as
\be K'(u)[S]=\left.\frac {\part }{\part \varepsilon }\right|
_{\varepsilon=0}K(u+\varepsilon S),\ \Phi'(u)[S]=
\left.\frac {\part }{\part \varepsilon }\right|
_{\varepsilon=0}\Phi(u+\varepsilon S).
\ee
\end{defi}

We recall that the commutator between two vector functions 
$K,S\in {\cal A}^q$ is given as
\be [K,S]=K'(u)[S]-S'(u)[K].\label{commutator}\ee 
The space ${\cal A}^q$ constitutes a Lie algebra under the bilinear operation
(\ref{commutator}).

\begin{defi}
A linear operator $\Phi(u): {\cal A}^q \to {\cal A}^q$ is called a 
hereditary symmetry  operator \cite{Fuchssteiner1979} 
if it satisfies the following condition
\be \Phi'(u)[\Phi K]S-\Phi '(u)[\Phi S]K-\Phi \{ \Phi '(u)[K]S
-\Phi'(u)[S]K\}=0 \label{hereditaryp}\ee
for arbitrary vector functions $K,S\in {\cal A}^q.$ 
\end{defi}  

An equivalent definition of a hereditary symmetry  operator $\Phi(u)
 :{\cal A}^q\to {\cal A}^q$ is that besides the linearity of $\Phi(u)$, 
its Nijenhuis torsion \cite{Kosmann-Schwarzbach} \cite{Thompson} 
$N_\Phi (K,S)$ vanishes for all $K,S\in {\cal A}^q$, i.e.
\bea  
N_\Phi (K,S)&:=&[\Phi K,\Phi S]-\Phi [\Phi K, S]-\Phi [K,\Phi S]
+\Phi ^2[K,S]
\nonumber\\ &=&(L_{\Phi S}\Phi)K-
\Phi (L_S\Phi) K=0,\label{Nijenhuisp}
\eea
where 
a Lie derivative $L_K\Phi$ of $\Phi(u): {\cal A}^q \to {\cal A}^q$ 
with respect to $K\in {\cal A}^q $ is given by
\be L_K\Phi=\Phi '[K]-[K',\Phi], \ee
or more precisely, 
\be (L_K\Phi)S=\Phi '(u)[K]S-K'(u)[\Phi S]+\Phi K'(u)[S],\ S\in 
{\cal A}^q.\ee

If a hereditary symmetry  operator $\Phi(u)$
has a zero Lie derivative $L_K\Phi=0$ with respect to 
$K\in {\cal A}^q$, then 
we have (for example, see  \cite{Fuchssteiner1979} \cite{Ma1990}) 
\be [\Phi ^mK,\Phi ^nK]=0,\ m,n\ge 0.\ee
Therefore each system of evolution equations
among the hierarchy 
\be u_t=\Phi ^nK,\ n\ge 0,\ee
has infinitely many commutative symmetries $\Phi^mK,\, m\ge 0.$
Such a vector field $K\in {\cal A}^q$ may often be chosen as $u_x$,
which will be seen later on.

The next section of the paper will examine two models of candidates for 
hereditary symmetry  operators. 
It will then go on to exhibit concrete examples 
of the general cases 
established in the second section. Finally, the fourth section 
will provide us with a summary and some concluding remarks.

\section{Extending hereditary symmetry  operators}
\setcounter{equation}{0}

Let us assume that 
\bea && u_k=(u_k^1,\cdots, u_k^q)^T,\  1\le k\le N,\nonumber \\ &&
u=(u_1^T,\cdots,u_N^T)^T=
(u_1^1,\cdots,u_1^q,\cdots, u_N^1,\cdots, u_N^q)^T.\nonumber \eea 
Throughout this paper, we need the following condition 
\be \Phi _k'(u_k)=\Phi _l'(u_l), \ 1\le k,l\le N, \label{linearp}\ee 
for a set of operators
$\Phi_k(u_k):\ {\cal A}^q(u_k)\to {\cal A}^q(u_k),\ 1\le k\le N$.
This reflects a kind of linearity property of the operators with
respect to the dependent variables $u_k,\, 1\le k\le N$.
We point out that there do exist such sets of operators $\Phi_k(u_k)$.
Some examples will be given in the next section.

Let us consider the first form of candidates for hereditary symmetry  operators
 \be \Phi (u)=\left(
\sum_{k=1}^Nc_{ij}^k\Phi _k(u_k) \right)_{N\times N}, \label{form1}\ee
where $\{c_{ij}^k|\, i,j,k=1,2,\cdots N\}$ is a set of given constants. 
Apparently we can define a linear operator
\[\Phi(u): {\underbrace{{\cal A}^q(u)\times \cdots \times 
{\cal A}^q(u)}_{N}}\to 
{\underbrace{{\cal A}^q(u)\times \cdots \times {\cal A}^q(u)}_{N}}\, ,\]
where a vector function of ${\cal A}^{q}(u)$ depends on 
all the dependent variables $u_1,\cdots,u_N$,
not just certain dependent variable $u_k$. 

\begin{thm} \label{model1}
(i)
If all $\Phi_k(u_k):{\cal A}^{q}(u_k)\to {\cal A}^{q}(u_k),
\ 1\le k\le N,$ are hereditary symmetry  operators satisfying 
the linearity 
condition (\ref{linearp}) and the constants $c_{ij}^k,\, 1\le i,j,k\le N$,
 satisfy the following coupled condition
\be 
\sum_{k=1}^Nc_{ij}^kc_{kn}^l=\sum_{k=1}^Nc_{ik}^lc_{kj}^n=
\sum_{k=1}^Nc_{in}^kc_{kj}^l,\ 
1\le i,j,l,n\le N,\label{realcond}\ee
then the operator $\Phi(u):{\cal A}^{Nq}(u)\to {\cal A}^{Nq}(u)$ 
defined by (\ref{form1}) is a hereditary symmetry  operator. 

(ii) If $L_{u_{kx}}\Phi_k=0$ for all $\Phi _k(u_k),
\ 1\le k\le N,$ then $L_{u_x}\Phi=0.$
\end{thm}

\noindent {\bf Proof:}
We only need to prove that $\Phi (u)$ satisfies the hereditary condition
(\ref{hereditaryp}), because the proof of the rest requirements is obvious.
Noting that ${\cal A}^q(u)$ is composed of column vector functions,
we may assume for $K,S\in {\cal A}^{Nq}(u)$ that 
\[K=(K_1^T,\cdots , K_N^T)^T,\ S=(S_1^T,\cdots , S_N^T)^T,
\ K_i,S_i\in {\cal A}^{q}(u),\ 1\le i\le N ,\]
and we often need to write $(X)_i=X_i,\ 1\le i\le N,$ when a vector function
$X\in {\cal A}^{Nq}(u)$ itself is complicated.
In this way we have 
\bea && 
\Phi K=((\Phi K)_1^T,\cdots,(\Phi K)_N^T)^T, \ (\Phi K)_i
=\sum_{l,n=1}^Nc_{in}^l\Phi_l(u_l)K_n ,\ 1\le i\le N,\nonumber \\
&& \Phi'(u)[\Phi K]=\left(\sum_{k=1}^Nc_{ij}^k\Phi_k'(u_k)\Bigl [
\sum_{l,n=1}^Nc_{kn}^l\Phi_l(u_l)K_n \Bigr ]\right)_{N\times N},
\nonumber \\
&& (\Phi'(u)[\Phi K]S)_i=
\sum_{j,k,l,n=1}^Nc_{ij}^kc_{kn}^l\Phi_k'(u_k)[\Phi_l
(u_l)K_n]S_j,\ 1\le i\le N,\nonumber \\ &&
 (\Phi\Phi'(u)[K]S)_i=
 \sum_{j,k,l,n=1}^Nc_{ik}^lc_{kj}^n\Phi_l(u_l)
\Phi_n'(u_n)[K_n]S_j,\ 1\le i\le  N.\nonumber \eea
Therefore by the linearity condition (\ref{linearp}), we can obtain 
\bea && 
(\Phi'(u)[\Phi K]S-\Phi '(u)[\Phi S]K-\Phi \{ \Phi '(u)[K]S
-\Phi'(u)[S]K\})_i\nonumber \\&=&
\sum_{j,l,n=1}^Nf(i,j,l,n)\bigl\{\bigr.
\Phi'_l(u_l)[\Phi _l(u_l)K_n]S_j-\Phi '_l(u_l)
[\Phi_l(u_l) S_j]K_n \nonumber \\&& 
-\Phi _l(u_l)\{ \Phi_l '(u_l)[K_n]S_j
-\Phi'_l(u_l)[S_j]K_n\}\bigl.\bigr \},\ 1\le i,j,n,l\le N,\label{last} \eea
where $f(i,j,l,n)$ is given by 
\[f(i,j,l,n):=\sum_{k=1}^Nc_{ij}^kc_{kn}^l=\sum_{k=1}^Nc_{ik}^lc_{kj}^n=
\sum_{k=1}^Nc_{in}^kc_{kj}^l=\sum_{k=1}^Nc_{ik}^lc_{kn}^j ,\ 
1\le i,j,l,n\le N.\]
This is well defined due to (\ref{realcond}). Actually the last equality
above may be obtained by changing two indices $n,j$ in the first equality
of (\ref{realcond}). 
Each term in the right side of (\ref{last}) 
is equal to zero because of the hereditary property of 
$\Phi_l(u_l),\,1\le l\le N,$ and thus $\Phi (u)$ satisfies the hereditary 
condition (\ref{hereditaryp}), indeed.
The proof is completed.
$\vrule width 1mm height 3mm depth 0mm$

Let us now 
consider the second form of candidates for hereditary symmetry  operators
\begin{equation}\label{form2}
 \Phi (u)=\left[\begin{array}{cccc}
0&\cdots&0& \Phi_1(u_1)\\
E_q&\cdots&0&\Phi_2(u_2)\\
\vdots & \ddots &\vdots & \vdots\\
0&\cdots& E_q&\Phi_N(u_N)\end{array}\right],
\end{equation}
where the matrix 
$E_q$ is the unit matrix of order $q$, i.e.
$E_q=\textrm{diag}\underbrace{(1,\cdots,1)}_q$. 

\begin{thm} \label{model2}
(i) If the operators $\Phi_k(u_k):{\cal A}^{q}(u_k)\to{\cal A}^{q}(u_k) 
,\,1\le k\le N,$ satisfy the linearity condition (\ref{linearp}), 
then the operator $\Phi(u):{\cal A}^{Nq}(u)\to{\cal A}^{Nq}(u) $
defined by (\ref{form2}) is hereditary if and only if the operators 
$\Phi_k(u_k),\, 1\le k\le N,$ are all hereditary.

(ii) The condition $L_{u_x}\Phi=0$ holds if and only if all the conditions
 $L_{u_{kx}}\Phi_k=0,\ 1\le k\le N,$ hold. 
\end{thm}

\noindent {\bf Proof:}
Similarly noting that ${\cal A}^q(u)$ is composed of column vector functions,
we may make the same assumption for $K,S\in {\cal A}^{Nq}(u)$:
\[K=(K_1^T,\cdots , K_N^T)^T,\ S=(S_1^T,\cdots , S_N^T)^T,
\ K_i,S_i\in {\cal A}^{q}(u),\ 1\le i\le N.\]  
Then we can obtain 
\bea && 
\Phi'(u)[\Phi K]S=\left [ \ba {c}\Phi_1'(u_1)[\Phi_1K_N]S_N\\
\Phi_2'(u_2)[K_1+\Phi_2K_N]S_N\\ \vdots \\
\Phi_N'(u_N)[K_{N-1}+\Phi_NK_N]S_N
 \ea \right],
\nonumber \\ && 
\Phi \Phi'(u)[ K]S=\left [ \ba {c}\Phi_1\Phi_N'(u_N)[K_N]S_N\\
\Phi_1'(u_1)[K_1]S_N+\Phi_2\Phi_N'(u_N)[K_N]S_N\\ \vdots \\
\Phi_{N-1}'(u_{N-1})[K_{N-1}]S_N+\Phi_N\Phi_N'(u_N)[K_N]S_N
 \ea \right],
\nonumber \\ &&
L_{u_x}\Phi =\left[ \ba {cccc} 
0 &\cdots &0 & \Phi_1'(u_1)[u_{1x}]-(\part \Phi_1-\Phi_1\part )\\
\vdots & &\vdots & \vdots \\
0 &\cdots &0 & \Phi_N'(u_N)[u_{Nx}]-(\part \Phi_N-\Phi_N\part )
\ea \right].
\nonumber\eea
Based upon the above three equalities and the linearity condition 
(\ref{linearp}), we can easily obtain the required 
results. So the proof is finished.  
$\vrule width 1mm height 3mm depth 0mm$

\section{Concrete examples}
\setcounter{equation}{0}

Basic scalar hereditary symmetry  
operators satisfying the linearity condition 
(\ref{linearp}) can be one of the following two sets
\be \Phi_i(u_i)=\alpha _i +\beta _i \part ^2+\gamma (\part u_i\part ^{-1}
+u_i),\ 1\le i\le N, \ee 
\be \Phi_i(u_i)=\alpha _i\part  +\gamma ( u_{ix}\part ^{-1}
+u_i),\ 1\le i\le N, \label{secondset}\ee
where $\part =\part /\part x$ and 
$\alpha _i,\beta _i,\gamma  $ are arbitrary constants. 
Of course, matrix hereditary symmetry  operators 
satisfying the linearity condition (\ref{linearp}) may be 
chosen and some of such examples have been given in Refs.
\cite{Ma1993,Ma1993b,Liu,MaZ}. Later on we will see 
two special examples while discussing extension problems. 
On the other hand, such sets of 
hereditary symmetry  operators may be 
generated directly from the above operators by Theorem \ref{model1}
and Theorem \ref{model2} in the previous 
section or by perturbation around solutions as in Refs. 
\cite{MaF1996a} \cite{MaF1996b}.     
Note that all the above hereditary symmetry  
operators satisfy $L_{u_{ix}}\Phi_i=0,
\ 1\le i\le N$.
Therefore among the corresponding hierarchy $u_t=\Phi ^nu_x,\,n\ge 0$, 
each system of evolution equations 
has infinitely many commutative symmetries, 
because we have $[\Phi ^mu_x,\Phi^nu_x]=0$ if $\Phi(u)$ is hereditary.

\subsection{Hereditary symmetry  operators of the first form:}

{\bf Example 1:} Let us choose 
\be c_{ij}^k=f(i)g(j)g(k),\ 1\le i,j,k\le N,\label{trivialset}\ee
where  $f,g$ may be arbitrary functions.  
The set of constants $\{c_{ij}^k\}$ satisfies the coupled
condition (\ref{realcond})
and thus the corresponding 
operator $\Phi(u)$ defined by (\ref{form1}) is hereditary
if each $\Phi _k(u_k)$ is hereditary and the linearity condition 
(\ref{linearp}) holds.
In particular, upon choosing $f(1)=g(1)=1,\,g(2)=2,\,f(2)=-3$,
 we have the following special hereditary symmetry  operator 
\[ \Phi(u)=\left [
 \ba {cc} \Phi_1(u_1)+2\Phi_2(u_2)& 2\Phi_1(u_1)+4\Phi_2(u_2)\\
-3\Phi_1(u_1)-6\Phi_2(u_2)& -6\Phi_1(u_1)-12\Phi_2(u_2)\ea
\right ],\]
where we require that $\Phi_1(u_1)$ and $\Phi_2(u_2)$ are hereditary and 
that $\Phi_1'(u_1)=\Phi_2'(u_2)$.
The second row of this operator 
is obtained by multiplying the first row by a constant $-3$
and so the operator is trivial. Due to the same fact, all hereditary
symmetry operators resulted from (\ref{trivialset}) are trivial.

Let us now choose 
\be c_{ij}^k=\delta _{kl},\ l=i+j-p\ (\textrm{mod}\, N) ,\ee
where $1\le p\le N$ is fixed and $\delta_{kl}$ 
denotes the Kronecker symbol again. The corresponding operators defined by 
(\ref{form1}) becomes
\be \Phi(u)=\left [ \ba {cccc}
\Phi _{2-p}(u_{2-p}) &\Phi _{1-p}(u_{1-p}) & \cdots & \Phi _{N-p+1}(u_{N-p+1})
\vspace{2mm}\\
\Phi _{3-p}(u_{3-p}) &\Phi _{2-p}(u_{2-p}) & \cdots & \Phi _{N-p+2}(u_{N-p+2})
\vspace{2mm}\\
\vdots &\vdots &\ddots & \vdots \vspace{2mm}\\
\Phi _{N-p+1}(u_{N-p+1}) &\Phi _{N-p+2}(u_{N-p+2}) & \cdots & 
\Phi _{2N-p}(u_{2N-p})
\ea \right] ,\label{phiform1}\ee
where we need to use $\Phi_i(u_i)=\Phi_j(u_j)$ if $i=j\ 
(\textrm{mod}\,N)$ to determine the operators involved, for example, 
$\Phi_{2-p}(u_{2-p})=\Phi _{N}(u_N)$ when $p=2$.

It can be proved that the coupled condition 
(\ref{realcond}) requires $N=2$. Thus among the above operators,
we have only two candidates of 
hereditary symmetry  operators satisfying (\ref{realcond})
\be \Phi(u)= \left [ \ba {cc} \Phi_1(u_1)&\Phi_2(u_2)\vspace{2mm}\\
\Phi_2(u_2) &\Phi_1(u_1) \ea \right ],\ 
 \Phi(u)= \left [ \ba {cc} \Phi_2(u_2)&\Phi_1(u_1)\vspace{2mm}\\
\Phi_1(u_1) &\Phi_2(u_2) \ea \right ],\ u=\left [\ba {c} 
u_1\vspace{2mm}\\
u_2 \ea \right ]. 
\ee  
Note that here $u_1$ and $u_2$ may be vector functions.
These two operators are symmetric and thus they can be 
diagonalizable. Actually 
they can be diagonalized by a linear transformation of the potentials $u_1$
and $u_2$. Therefore they are also trivial. What we show above is
that there is no interesting hereditary symmetry operator among the operators
defined by (\ref{phiform1}). 

{\bf Example 2:} Let us choose
\be c_{ij}^k=\delta _{kl},\ l=i-j+p\ (\textrm{mod}\, N), \ee
where $1\le p\le N$ is also fixed and $\delta_{kl}$ still
denotes the Kronecker symbol. In this case, we have 
\bea && 
\sum_{k=1}^Nc_{ij}^kc_{kn}^l=\sum_{k=1}^Nc_{ik}^lc_{kj}^n=
\sum_{k=1}^Nc_{in}^kc_{kj}^l
\nonumber \\
& =&
\left \{ \ba {cl} 
1 &\quad \textrm{when} \ i-j-n-l+2p=0 
\ (\textrm{mod}\, N),\vspace{2mm}\\
0 &\quad \textrm{otherwise} ,
\ea \right .\nonumber
 \eea 
which implies that the coupled condition 
 (\ref{realcond}) automatically holds.
Thus we have a set of candidates for 
hereditary symmetry  operators
\be \Phi(u)= \left [ \ba {ccccccc} \Phi_p(u_p)&\Phi_{p-1}(u_{p-1})&\cdots &
\Phi_1(u_1)&\Phi _N(u_N)& \cdots &\Phi _{p+1}(u_{p+1})\vspace{3mm}\\
\Phi _{p+1}(u_{p+1})& \Phi_p(u_p)& \ddots &\ddots &\Phi_1(u_1) 
&\ddots &\vdots \vspace{3mm} \\
\vdots & \ddots &\ddots &\ddots &\ddots &\ddots &   \Phi_N(u_N)\vspace{3mm} \\
\Phi _N(u_N) & &\ddots &\ddots &\ddots &\ddots & \Phi_1(u_1)\vspace{3mm}\\
 \Phi_1(u_1)&\ddots & &\ddots &\ddots  &\ddots &\vdots\vspace{3mm}\\
\vdots &\ddots &\Phi_N(u_N) & &\ddots & \Phi_p(u_p) &\Phi_{p-1}(u_{p-1})
\vspace{3mm} \\
\Phi_{p-1}(u_{p-1}) &\cdots &\Phi_1(u_1) &\Phi_N(u_N)& \cdots & \Phi_{p+1}
(u_{p+1})&\Phi_p(u_p) \ea \right ],\ee
where  we also need to use $\Phi_i(u_i)=\Phi_j(u_j)$ if $i=j\ 
(\textrm{mod}\,N)$ to determine the operators involved.
In particular, we can obtain a candidate of hereditary symmetry  operators
 \be \Phi(u)= \left [ \ba {cccc} \Phi_1(u_1)&\Phi_N(u_N)&\cdots &\Phi_2(u_2)
\vspace{1mm}\\
\Phi_2(u_2) &\Phi_1(u_1)& \ddots &\vdots \vspace{1mm}\\
\vdots & \vdots &\ddots &\Phi_N(u_N) \vspace{1mm}\\
\Phi_N(u_N)& \Phi_{N-1}(u_{N-1})& \cdots &\Phi_1(u_1)
 \ea \right ]. 
\ee  

The $N=3$ case of the above operator with the scalar operators
\[\Phi_i(u_i)=\beta _i \part ^2 +
(\part u_i \part ^{-1} +u_i),\ 1\le i\le 3,\] 
gives a hierarchy of nonlinear systems $u_{t}=(\Phi(u))^nu_x,\ n\ge 1$,
among which the first nonlinear system reads as
\be 
\left\{ \ba {c} u_{1t}=
\beta_1u_{1xxx}+\beta_3u_{2xxx}+\beta_2u_{3xxx}+ 3u_1u_{1x}+3(u_2u_{3})_x,
\vspace{2mm} \\ u_{2t}=
\beta_2u_{1xxx}+\beta_1u_{2xxx}+\beta_3u_{3xxx}+ 3u_3u_{3x}+ 3(u_1u_{2})_x,
\vspace{2mm}
 \\ u_{3t}=
\beta_3u_{1xxx}+\beta_2u_{2xxx}+\beta_1u_{3xxx}+3u_2u_{2x}+ 3(u_1u_{3})_x.
 \ea \right.
\label{3x3system}\ee
This system is not symmetric with respect to $u_1,u_2,u_3$, and generally it
can not be separated under a real linear transformation of the potentials
$u_1,u_2,u_3$. One of the reasons is that the matrix
\[A=\left[ \ba {ccc} \beta _1 & \beta _3 & \beta _2 \vspace{2mm}\\
 \beta _2 & \beta _1 & \beta _3 \vspace{2mm}\\
 \beta _3 & \beta _2 & \beta _1
\ea \right]\]
can not be always diagonalized for all values of 
$\beta _1,\, \beta_2,\, \beta _3$.
When $u_1=u_2=u_3$, the system is reduced to 
the KdV equation up to a constant coefficient.
It also provides an example of the general systems discussed by G\"urses 
et al. in Ref. \cite{GursesK}. 

{\bf Example 3:}
If we choose 
\be c_{ij}^k=\delta _{i-j,k-p},\ee
where  $p$ is an integer and $\delta_{kl}$ denotes the Kronecker symbol.
For two cases of $2-N\le p\le 1$ and $N\le p\le 2N-1$,
the coupled condition (\ref{realcond}) can be satisfied, 
because we have
\bea && 
\sum_{k=1}^Nc_{ij}^kc_{kn}^l=\sum_{k=1}^Nc_{ik}^lc_{kj}^n=
\sum_{k=1}^Nc_{in}^kc_{kj}^l
\nonumber \\
& =&
\left \{ \ba {cl} 
1 &\quad \textrm{when} \ i-j-n-l+2p=0,\vspace{2mm}\\
0 &\quad \textrm{otherwise} .
\ea \right .\nonumber
 \eea 
We should note in proving the above equality that we have 
\[ 1\le i-j+p=n+l-p\le N, \ 1\le i-l+p=n+j-p\le N,\ 1\le i-n+p=j+l-p\le N, \]
when $i-j-n-l+2p=0$.
But for the case of $1<p<N$, upon choosing $i=n=N, \,j=p+1,\,l=p-1$,
we have
\[ \sum_{k=1}^Nc_{ij}^kc_{kn}^l=1,\ \sum_{k=1}^Nc_{ik}^lc_{kj}^n=0,\]
and thus the coupled condition (\ref{realcond}) can not be satisfied.

Note that when $p<2-N$ or $p>2N-1$, the resulting operators are all 
zero operators.
Therefore we can obtain
only two sets of candidates for hereditary symmetry  operators
\bea  \Phi(u)&=& \left [ \ba {cccc} \Phi_p(u_p) & & & 0\\
\Phi_{p+1}(u_{p+1})&\ddots & & \\
\vdots & \ddots&\ddots & \\
 \Phi_{p+N-1}(u_{p+N-1})& \cdots & \Phi_{p+1}(u_{p+1}) & \Phi_p(u_p)
\ea \right ],\ 2-N\le p\le 1,
\label{thirdho1} \vspace{2mm}
\\ 
\Phi(u)&=& \left [ \ba {cccc} \Phi_p(u_p) & \Phi_{p-1}(u_{p-1}) &\cdots & 
\Phi_{p-N+1}(u_{p-N+1})\\
& \ddots&\ddots & \vdots \\
& &\ddots &\Phi_{p-1}(u_{p-1})\\
 0& & & \Phi_{p}(u_{p})
\ea \right ],\ N\le p\le 2N-1,
\label{thirdho2} \quad \eea
where we accept that $\Phi_i(u_i)=0$ if $i\le 0$ or $i\ge N+1$. 
These two sets of operators can be linked by a transformation
$(u_1,u_2,\cdots,u_N)\leftrightarrow (u_N,u_{N-1},\cdots,u_1)$. 

When we take 
\[\Phi_i(u_i)=\alpha _i \part ^2+2(\part u _{i}\part ^{-1}+u_i), \ 
1\le i\le N,\]
where $\alpha _i,\,1\le i\le N$, are arbitrary constants,
as basic hereditary symmetry  operators,
we obtain $N$ hierarchies of nonlinear systems of KdV type starting from 
the operators in (\ref{thirdho1}).
A special choice with $\alpha _1=1,\ \alpha_i=0,\, 2\le i\le N$,
and $p=1$ leads to  
the perturbation systems of the KdV equation
generated from perturbation around solutions
in Ref. \cite{MaF1996a}. 
Another special choice with $N=2$ and $p=1$ leads to 
the following system
\be \left\{ 
\ba {l} u_{1t}=\alpha _1 u_{1xxx}+6u_1u_{1x}, \vspace{2mm}\\
u_{2t}=\alpha _2u_{1xxx}+\alpha _1u_{2xxx}+6(u_1u_2)_x.
 \ea \right.\ee

We can also choose a pair of hereditary symmetry operators
in Ref. \cite{Ma1993c}
\be  \Phi_1(u_1)=\left [\ba {cc} 
u_{1x}^1 \part ^{-1}+2u_1^1 & u_1^2 +\alpha  \part 
\vspace{2mm}\\
u_{1x}^2\part ^{-1} +u_1^2 -\alpha  \part &0
\ea \right], \ \Phi _2(u_2)=
\left [\ba {cc} 
u_{2x}^1 \part ^{-1}+2u_2^1 & u_2^2 +\beta \part 
\vspace{2mm}\\
u_{2x}^2\part ^{-1} +u_2^2 -\beta \part &0
\ea \right]\label{heredma1} \ee
as basic hereditary symmetry  operators
with $u_1=(u_1^1,u_1^2)^T$ and $u_2=(u_2^1,u_2^2)^T$ and two 
arbitrary constants $\alpha$ and $\beta$. 
Then we can obtain a $4\times 4$ matrix hereditary symmetry  operator
 \be \Phi (u)=\left [ \ba {cccc}
u_{1x} \part ^{-1}+2u_1 & u_2 +\alpha \part &0 &0
\vspace{2mm}\\
u_{2x}\part ^{-1} +u_2 -\alpha  \part &0 &0&0
\vspace{2mm}\\ 
u_{3x} \part ^{-1}+2u_3 & u_4 +\beta \part & 
u_{1x} \part ^{-1}+2u_1 & u_2 +\alpha  \part
\vspace{2mm}\\
u_{4x}\part ^{-1} +u_4 -\beta \part &0&
u_{2x}\part ^{-1} +u_2 -\alpha  \part &0
\ea \right] ,\ u=\left[\ba {c} u_1\vspace{2mm}\\u_2\vspace{2mm}\\
u_3\vspace{2mm}\\u_4\ea \right] , 
\label{4x4operator1}\ee
with two arbitrary constants $\alpha $ and $\beta $.
Note that we rename the 
dependent variables $u^1_1,u_1^2,u_2^1,u_2^2$ as $u_1,u_2,u_3,u_4$, 
respectively.
The first nonlinear system in the hierarchy $u_t=(\Phi (u))^nu_x,\ n\ge 1$,
is the following
\bea \left\{ \ba {l} 
u_{1t}=\alpha  u_{2xx}+3u_1u_{1x}+u_2u_{2x},
 \vspace{2mm}\\
u_{2t}=-\alpha  u_{1xx}+(u_1u_2)_{x},
\vspace{2mm}\\
u_{3t}=\beta u_{2xx}
+\alpha  u_{4xx}+3(u_1u_{3})_x+(u_2u_4)_x
,\vspace{2mm}\\
u_{4t}=-\beta u_{1xx}-\alpha  u_{3xx}+(u_1u_4)_{x}+(u_2u_3)_{x}.
\ea  \right.\eea
This is of different type from one 
discussed in Ref. \cite{Svinolupov1989} because of the terms of the
second derivatives of potentials.

\subsection{Hereditary symmetry  operators of the second form:}

{\bf Example 4:} 
Let $\Phi(u)$ be defined by ({\ref{form2}).
The first nontrivial candidate of integrable systems among the hierarchy 
$u_t=(\Phi(u))^nu_x,\, n\ge 0,$
reads as
\be u_t=\left [\ba {c} u_1\vspace{2mm}\\
u_2\vspace{2mm}\\
\vdots \vspace{2mm}\\
u_N \ea \right]_t=\left [ \ba {c} 
\Phi_1(u_1)u_{Nx}\vspace{2mm}\\
u_{1x}+\Phi_2(u_2)u_{Nx}\vspace{2mm}\\
\vdots\vspace{2mm}\\
u_{N-1,x}+\Phi_N(u_N)u_{Nx}
\ea \right]. \ee 
If we choose the basic scalar hereditary symmetry  operators as follows
\[\Phi_i(u_i)=-\frac 14 \part ^2+(\part u_i\part ^{-1}+u_i),\ 1\le i\le N,\]
then the corresponding hereditary symmetry  operator $\Phi(u) $ determined by 
(\ref{form2}) becomes 
\be \Phi (u)=\left[\ba {cccc}
0&\cdots&0&  -\frac 14 \part ^2+(\part u_1\part ^{-1}+u_1)\vspace{2mm}
\\ 1&\cdots&0&-\frac 14 \part ^2+(\part u_2\part ^{-1}+u_2)\vspace{2mm}\\
 \vdots& \ddots & \vdots& \vdots\vspace{2mm}\\
0&\cdots& 1&-\frac 14 \part ^2+(\part u_N\part ^{-1}+u_N)
  \ea \right]. \ee
This generates the coupled KdV systems \cite{Ma1993} \cite{FordyA}.

If we choose the basic scalar hereditary symmetry  operators defined by 
(\ref{secondset}), then the  
corresponding hereditary symmetry operator
contains all hereditary symmetry operators appeared in Refs. 
\cite{Ma1993b,Liu,MaZ}. A special example gives a hereditary symmetry 
operator  
\be \Phi (u)=\left[\ba {ccccc}0&0&0&0&\alpha _1 \part +u_{1x}\part ^{-1}+u_1
\vspace{2mm} \\1&0&0&0&\alpha _2 \part +u_{2x}\part ^{-1}+u_2
\vspace{2mm} \\0&1&0&0&\alpha _3\part +u_{3x}\part ^{-1}+u_3
\vspace{2mm} \\0&0&1&0&\alpha _4 \part +u_{4x}\part ^{-1}+u_4
\vspace{2mm} \\0&0&0&1&\alpha _5\part +u_{5x}\part ^{-1}+u_5
\ea \right],\ 
u=\left[\ba {c} u_1\vspace{2mm} \\ u_2\vspace{2mm} \\
u_3\vspace{2mm} \\u_4\vspace{2mm} \\ u_5  \ea \right],
 \ee 
and a nonlinear system 
\be \left \{\ba {l} u_{1t}= \alpha _1u_{5xx}+(u_1u_5)_x,
\vspace{2mm}\\ u_{2t}= u_{1x}+\alpha _2u_{5xx}+(u_2u_5)_x,
\vspace{2mm}\\u_{3t}= u_{2x}+\alpha _3u_{5xx}+(u_3u_5)_x,
\vspace{2mm}\\u_{4t}= u_{3x}+\alpha _4u_{5xx}+(u_4u_5)_x,
\vspace{2mm}\\u_{5t}= u_{4x}+\alpha _5u_{5xx}+2u_5u_{5x},
\ea \right .\ee
with five arbitrary constants $\alpha _i,\ 1\le i\le 5.$

{\bf Example 5:} 
Let us choose another pair of $2\times 2$ matrix operators 
\[ \Phi_1(u_1)=\left [\ba {cc} 
0&\beta _1\part +\gamma (u_{1x}^1\part 
^{-1}+u_1^1)\vspace{2mm}\\
\alpha _1&\beta _2\part +\gamma (u_{1x}^2\part 
^{-1}+u_1^2)
\ea \right], \ \Phi _2(u_2)=
\left [\ba {cc} 
0&\beta _3\part +\gamma (u_{2x}^1\part 
^{-1}+u_2^1)\vspace{2mm}\\
\alpha _2&\beta _4\part +\gamma (u_{2x}^2\part 
^{-1}+u_2^2)
\ea \right]\]
as basic hereditary symmetry  operators with 
$u_1= (u_1^1,u_1^2)^T$ and $u_2=(u_2^1,u_2^2)^T$.
Then by Theorem \ref{model2},
we obtain a $4\times 4$ matrix hereditary symmetry  operator
 \be \Phi (u)=\left [ \ba {cccc} 0& 0& 0& \beta _1\part +\gamma (u_{1x}\part 
^{-1}+u_1)\vspace{2mm}\\
0&0&\alpha _1&\beta _2\part +\gamma (u_{2x}\part 
^{-1}+u_2)\vspace{2mm}\\ 1&0&0&\beta _3\part +\gamma (u_{3x}\part 
^{-1}+u_3)\vspace{2mm}\\0&1&\alpha_2&\beta _4\part +\gamma (u_{4x}\part 
^{-1}+u_4)
\ea \right] ,\ u=\left[\ba {c} u_1\vspace{2mm}\\u_2\vspace{2mm}\\
u_3\vspace{2mm}\\u_4\ea \right] . \label{4x4}
\ee
where $\alpha_i,\beta_i,\gamma$ are arbitrary constants and we rename the 
dependent variables $u^1_1,u_1^2,u_2^1,u_2^2$ as $u_1,u_2,u_3,u_4$, 
respectively.
The first nonlinear system from the corresponding hierarchy 
is the following
\be \left \{\ba {l} u_{1t}=\beta _1u_{4xx}+\gamma (u_1u_4)_x,\vspace{2mm}\\
u_{2t}=\alpha _1 u_{3x}+\beta _2u_{4xx}+\gamma (u_2u_4)_x,\vspace{2mm}
\\u_{3t}=u_{1x}+\beta _3u_{4xx}+\gamma (u_3u_4)_x,\vspace{2mm}\\u_{4t}=
u_{2x}+\alpha _2u_{3x}+\beta _4 u_{4xx}+2\gamma u_4u_{4x}.
\ea \right .\label{4x4system}\ee
This system is reduced to the Burgers equation up to a constant coefficient,
if we make a special choice  
\[u_1=u_2=\alpha _1=\alpha _2=\beta_1=\beta _2=0,\ u_3=u_4,\ \beta _3=
\beta _4.\] 

Let us next choose the following 
three $2\times 2$ matrix operators in Ref. \cite{Ma1993c}
\be  \Phi_i(u_i)=\left [\ba {cc} 
 u_i^2+\alpha _i \part & u_{ix}^1\part ^{-1}+2u_i^1
\vspace{2mm}\\0&
u _{ix}^2\part ^{-1}+u_i^2-\alpha _i\part 
\ea \right], \ 1\le i\le 3, 
\ee
as basic hereditary symmetry  operators 
with $u_i=(u_i^1,u_i^2)^T,\,1\le i\le 3$.
It is quite interesting to observe that the above hereditary symmetry operators
can be obtained by interchanging two columns 
of the hereditary symmetry operators in (\ref{heredma1}). 
Through Theorem \ref{model2},
we obtain a $6\times 6$ matrix hereditary symmetry  operator
 \be \Phi (u)=\left [ \ba {cccccc} 
0& 0&0&0& u_2+\alpha _1 \part &
u_{1x}\part ^{-1}+2u_1
\vspace{2mm}\\
0&0&0&0&0&
u _{2x}\part ^{-1}+u_2-\alpha _1\part 
\vspace{2mm}\\ 
1&0&0&0&u_4+\alpha _2 \part & u_{3x}\part ^{-1}+2u_3
\vspace{2mm}\\0&1&0&0&0&
u _{4x}\part ^{-1}+u_4-\alpha _2 \part
\vspace{2mm}\\
0&0&1&0& u_6+\alpha _3 \part & u_{5x}\part ^{-1}+2u_5
\vspace{2mm}\\0&0&0&1&0&
u _{6x}\part ^{-1}+u_6-\alpha _3 \part 
\ea \right] ,\ u=\left[\ba {c} u_1\vspace{2mm}\\u_2\vspace{2mm}\\
u_3\vspace{2mm}\\u_4\vspace{2mm}\\u_5\vspace{2mm}\\u_6\ea \right] ,
\label{6x6}
\ee
where $\alpha _i,\, 1\le i\le3,$ are arbitrary constants
and we rename the 
dependent variables $u^1_1,u_1^2,u_2^1,u_2^2,u_3^1,u_3^2$ 
as $u_1,u_2,u_3,u_4,u_5,u_6$, respectively.
The first nonlinear system of the corresponding hierarchy 
reads as
\be \left \{\ba {l} u_{1t}=
\alpha _1u_{5xx}+u_2u_{5x}+(u_1u_{6})_x+u_1u_{6x},\vspace{2mm}\\
u_{2t}=-\alpha _1u_{6xx}+(u_2u_6)_x,\vspace{2mm}
\\u_{3t}=u_{1x}+ \alpha _2 u_{5xx}+
 u_4u_{5x}+(u_3u_{6})_x+u_3u_{6x},\vspace{2mm}\\u_{4t}=
u_{2x}-\alpha _2 u_{6xx}+  (u_4u_{6})_x,
\vspace{2mm}
\\u_{5t}=u_{3x}+ \alpha _3 u_{5xx}+
 2(u_5u_{6})_x,\vspace{2mm}\\u_{6t}=
u_{4x}-\alpha _3 u_{6xx}+ 2 u_6u_{6x}.
\ea \right .\label{6x6system}\ee
This is another different example from the systems discussed by 
Svinolupov in \cite{Svinolupov1989}. 

\section{Conclusions and remarks}
\setcounter{equation}{0}

Two models of candidates for hereditary symmetry  operators are 
analyzed and 
some possible basic hereditary symmetry  operators are also given. 
Therefore according to the conditions in Theorem \ref{model1}
and Theorem \ref{model2},
many concrete nonlinear systems of evolution equations possessing 
infinitely many symmetries may be generated 
from various hereditary symmetry  
operators having a zero Lie derivative with respect
to $u_x$. Some particular cases are carefully discussed, along with
several corresponding nonlinear systems.

Our results provide a direct way to extend hereditary symmetry operators.
New resulting hereditary symmetry  operators,
for example, the  hereditary symmetry operators
shown in (\ref{4x4operator1}), (\ref{4x4}) and (\ref{6x6}), 
can also be chosen as 
basic ones satisfying the linearity condition (\ref{linearp}),
and then more complicated hereditary symmetry  operators can be generated 
by our idea of construction. Note that $x$ may be a vector 
and no condition has been 
imposed on the spatial dimension while examining two 
forms of candidates for hereditary symmetry  operators.
Therefore the idea is also valid for the case of high spatial dimensions,
which will be reported elsewhere.  
On the other hand, we hope that there will appear more concrete examples 
satisfying (\ref{realcond}) and more 
concrete models of hereditary symmetry  operators.

It is worthy pointing out that the coupled condition (\ref{realcond})
is only sufficient but not necessary. We may have counterexamples. For example,
a counterexample can be the following
\bea &&\Phi(u)=\left[\ba {cc} \Phi_1(u_1) &0 \vspace{2mm} \\ \Phi_2(u_2)+
a\Phi_1(u_1)&\Phi_1(u_1) \ea \right],\ 
u=\left[\ba {c} u_1\vspace{2mm}\\u_2\ea \right] ,
\vspace{2mm}\label{counterex1}\\
 &&\Phi_i(u_i)=\delta_{i1}\part ^2+
2(\part u_i\part ^{-1}+u_i),\ i=1,2, \label{counterex2}\eea
with an arbitrary non-zero constant $a$. 
In fact, for this operator $\Phi(u) $ we have
\[  (c_{ij}^1)= \left[\ba {cc}1&0\vspace{2mm} \\ a &1 \ea \right] ,
\  (c_{ij}^2)= \left[\ba {cc}0&0\vspace{2mm} \\ 1 &0 \ea \right] .
\]  
If we choose $i=l=n=1,\,j=2$, then we obtain
\[ \sum_{k=1}^2c_{ij}^kc_{kn}^l=0,\  \sum_{k=1}^2c_{ik}^lc_{kj}^n=a, \]
and thus the first equality in the coupled condition (\ref{realcond})
is not satisfied. But the operator $\Phi(u)$ defined by (\ref{counterex1})
and (\ref{counterex2}) is hereditary, which may be directly proved.

The coupled condition (\ref{realcond}) may also be viewed as a condition
on a finite-dimensional algebra 
with a basis $\textrm{e}_1,\,\textrm{e}_2,\,\cdots ,\textrm{e}_N$
and an operation
\be  \textrm{e}_i*\textrm{e}_j=\sum_{k=1}^Nc_{ij}^k\textrm{e}_k,\ 1\le i,j
\le N.\ee
But we have not yet known much about such kind 
of algebras.

\vskip 3mm
\noindent{\bf Acknowledgment:} The author
would like to thank M. Blaszak, B. Fuchssteiner, W. Oevel and
W. Strampp for useful 
discussions when he was visiting Germany,
and City University of Hong Kong for financial support.

\end{document}